%Paper: hep-th/9404093
%From: MICKELSSON@phcu.helsinki.fi
%Date: Fri, 15 Apr 1994 15:24:36 +0200 (EET)
%Date (revised): Fri, 03 Jun 1994 15:18:00 +0200 (EET)

\input amstex.tex
\documentstyle{amsppt}
\magnification\magstep1

%\redefine\Bbb{\bold}

\define\a{\alpha}

\define\g{\gamma}

\redefine\o{\omega}

\redefine\l{\lambda}
\redefine\L{\Lambda}

\define\gm{\bold g}

\define\<#1,#2>{\langle #1,#2\rangle}
\define\TR{\text{tr}}
\define\dep(#1,#2){\text{det}_{#1}#2}

\topmatter
\title WODZICKI RESIDUE AND ANOMALIES OF CURRENT ALGEBRAS \endtitle
\author Jouko Mickelsson \endauthor
\affil Royal Institute of Technology, Stockholm\\
e-mail jouko\@ theophys.kth.se, tel. 46 8 790 7278 \endaffil
\endtopmatter
\NoRunningHeads
\document
\baselineskip 18pt

ABSTRACT
The commutator anomalies (Schwinger terms) of current algebras in
$3+1$ dimensions are
computed in terms of the Wodzicki residue of pseudodifferential
operators; the result can be written as a (twisted) Radul 2-cocycle for the
Lie algebra of PSDO's. The construction of the (second
quantized) current algebra is closely related to
a geometric renormalization of the interaction Hamiltonian
$H_I=j_{\mu} A^{\mu}$ in gauge theory.

\vskip 0.6in
1. INTRODUCTION

\vskip 0.3in
One of the problems one meets all the time in quantum field theory is that
products of field operators at equal times in the same points in the
physical space are ill-defined. The first renormalization which one always
performs is the normal ordering of the operator products, i.e. a shift of
the energy lowering operators to the right, those giving a vanishing
contribution when acting on the Dirac vacuum. In fact, in  many cases in
1+1 dimensional models the normal ordering is sufficient to produce
well-defined second quantized operators (more precisely, operator
valued distributions). One such a case is the algebra of local charges
formed as certain quadratic expressions in field operators. In one
space dimension the construction leads to affine Kac-Moody and related
algebras.

In higher dimensions one needs further renormalizations in addition to the
normal ordering. The aim of this talk is to explain the renormalizations
needed for local charges in 3+1 space-time dimensions.  More specifically,
we shall study (chiral) fermions minimally coupled to external gauge fields.
The basic idea in the present renormalization scheme is to conjugate the
Gauss law generators by unitary operators, which are functions of the
external gauge field, in the one-particle space such that the resulting
conjugated operators can be quantized using the standard normal ordering
prescription. As a by-product, we obtain a new geometric renormalization
of the Dirac-Yang-Mills interaction hamiltonian.

In one space dimension there is a nontrivial 2-cocycle which defines a central
extension
of the Lie algebra of pseudodifferential operators, [KK]. This algebra can be
identified as the quantum $W_{\infty}$ algebra. It has as a subalgebra
(when the coefficients of the PSDO's are taken in a simple Lie algebra)
an affine algebra. The Kravchenko-Khesin cocycle has been generalized by
Radul to all  dimensions, [R].
We shall show that a twisted form of the Radul cocycle, when applied to the
renormalized
local charges, gives an extension of the naive current algebra which is
equal to the Mickelsson-Faddeev algebra, [M, F-Sh], [M2]. Thus the MF algebra
is closely related to a multidimensional version of the
$W_{\infty}$ algebra.

\vskip 0.3in

2. CENTRAL EXTENSION OF ALGEBRAS OF PSDO's

Let us first consider pseudodifferential operators in one dimension, on
a circle. Asymptotically, a  PSDO is a defined by a Laurent series
$$a(x,p) = \sum_{k\leq n} a_{k}(x) p^k\tag2.1$$
where $n$ is some integer and the $a_k$'s are smooth functions on the
circle. The momentum $p$ is the symbol of the operator $-i\partial_x.$
The product is defined as
$$a*b = \sum_{k=0,1,2,\dots}\frac{(-i)^k}{k!} \partial_p^k a(x,p)
\partial_x^k b(x,p). \tag2.2$$
Note that each $(a*b)_j$ is a finite sum of products of derivatives in the
coefficients $a_i,b_i.$

The algebra of PSDO's becomes a Lie algebra $\Cal B$ under the commutator
$[a,b]=a*b -b*a.$

The Adler-Manin residue of a PSDO is defined as
$$Res(a) = \frac{1}{2\pi}\int \TR\, a_{-1} dx\tag2.3$$
where we have included the trace in order to allow a generalization to
matrix valued PSDO's. The residue behaves like a trace on the algebra of
PSDO's. It is obviously a linear functional and furthermore it satisfies
$$Res([a,b])=0.\tag2.4$$

The function log$(p)$ is not a PSDO (since its expansion contains arbitrarily
high powers of $p$) but nevertheless one can define a Lie algebra 2-cocycle,
the KK cocycle, by the formula
$$c(a,b) = Res[log(p), a]*b.\tag2.5$$
This is because the commutator $[log(p),a]$ is a PSDO,
$$[log(p),a] = \sum_{k=1,2,\dots}- \frac{i^k}{k} \partial_x^k a(x,p) p^{-k}
\tag2.6$$
The 2-cocycle property
$$c(a,[b,c]) +\text{ cycl. permutations } =0$$
is a simple consequence of (2.4).

In the case when $a,b$ are zeroth order PSDO's, i.e. they are multiplication
operators, the value of the KK cocycle is
$$c(a,b) = \frac{i}{2\pi} \int \TR\, a(x) b'(x) dx.\tag2.7$$
This is exactly the central term of an affine Lie algebra (when $a,b$ take
values in a simple Lie algebra).

The KK cocycle can be defined in any number of space dimensions, [R]. Thus
one might wonder whether the higher dimensional Radul cocyles have anything to
do with anomalies of current algebras. The physically relevant extensions of
current algebras in higher dimensions are generally not central extensions
but extensions by some abelian ideal. For this reason it is not immediately
obvious what is the relevance of the Radul cocycle in higher dimensions.
We shall clarify this matter in section 5.

\vskip 0.3in
3. THE WODZICKI RESIDUE

\vskip 0.3in
Let $M$ be a compact manifold of dimension $n.$ A PSDO on $M$, with
coefficients
in a vector bundle $V$ over $M$, is locally given by a matrix valued symbol
$a(x,p).$
Here $x$ is a local coordinate on $M$ and $p$ is a fiber coordinate in the
cotangent bundle $T^*M.$  The product rule for symbols is determined from the
definition of the operator $A$ acting on sections of $V.$
$$(A\psi)(x) = \frac{1}{(2\pi)^{n/2} } \int e^{-ix\cdot p} a(x,p) \hat\psi(p)
d^n p\tag3.1$$
where $\hat \psi$ is the Fourier transform of the section $\psi.$ Thus the
symbol of the product $AB$ is
$$(a*b)(x,p) = \frac{1}{(2\pi)^n} \int e^{i(x-y)\cdot (p-q)} a(x,q)b(y,p)
d^n y d^n q. \tag3.2$$

The adjoint of $A$ (in the Hilbert space of square-integrable sections, the
measure defined by a Riemannian metric on $M$) is in general a complicated
expression in terms of the symbol $a.$ We shall give the formula only in the
euclidean case:
$$A^* \sim a^* + \Omega a^* + \frac{1}{2!}\Omega^2 a^* + \dots \tag3.3$$
where
$$\Omega = -i \sum_j \partial_{x_j} \partial_{p_j}$$
and $a^*$ is the matrix adjoint of the matrix valued symbol $a.$

Let $a_{-k}$ be a symbol of integral order $-k$ in $n$ space dimensions. Then
the
trace is asymptotically
$$\TR a_{-k}= \int d^nx \int d^n p \TR\, a_{-k}(x,p)\sim \int d^nx\int
|p|^{-k} |p|^{n-1} d|p|$$
from which follows that $a_{-k}$ is of trace-class if and only if $-k\leq
-n-1.$
Similarly, $p_{-k}$ is Hilbert-Schmidt if and only if the integer $-2k\leq
-n-1.$

Most of the time we are interested only on the asymptotic behaviour of the
symbols $a$ for large momenta $p.$ We assume that a PSDO has an asymptotic
expansion of the form
$$a = a_{k} +a_{k-1} +a_{k-2} \dots \tag 3.4$$
where each $a_k= a_k(x,p)$ is a smooth function of $x$ and of $p\neq 0$,
$a_k$ is homogeneous of degree $k$ in the momenta, $\frac{a_k}{|p|^{k+1}}
\to 0$ as $|p|\to \infty.$
One can show that the asymptotic expansion of the symbol $a*b$ can be
written as (compare with (2.2))
$$a*b = \sum_m \frac{(-i)^{|m|}}{m!} \left(\partial_{p_1}^{m_1} \dots
\partial_{p_n}^{m_n} a(x,p)\right) \left(\partial_{x_1}^{m_1} \dots
\partial_{x_n}^{m_n} b(x,p)\right)  \tag3.5$$
where the sum is over all multi-indices $m=(m_1,\dots,m_n),$ $|m|= m_1+
\dots + m_n,$ and $m!= m_1! \dots m_n!$ , [H].

The Wodzicki residue [W] of a PSDO $a$ is defined as a linear functional
which depends only on the component $a_{-n},$
$$Res(a) = \frac{1}{(2\pi)^n} \int_{|p|=1} \TR a_{-n}(x,p) \eta (d\eta)^{n-1}
\tag3.6$$
where $\eta = \sum p_k dx_k$ and $d\eta= \sum dp_k dx_k$ is the symplectic
2-form on the cotangent bundle $T^*M,$ $n>1.$ In the case $n=1$ this is
\it almost \rm  the
Adler-Manin residue, [Ad], [Ma]. The difference is the following. If $a_1=
\frac{\a(x)}{p}$ then the resudue is zero, because the unit sphere in momentum
space consists of two points $\pm 1$ and the momentum space integral is
$$\int_{|p|=1} \frac{\a(x)}{p} pdx = (\a(x)\vert_{p=+1} -\a(x)\vert_{p=-1})dx
=0.$$
However, if $a_{-1}=\frac{\a(x)}{|p|}$ then the integral becomes the sum
$\a(x) +\a(x) =2\a(x).$ Thus we can write
$$Res_{AM} (a) = \frac12 Res_W (\epsilon a)$$
where $\epsilon=p/|p|.$ Usually one redefines the Wodzicki residue in one
dimension so that it agrees precisely with the Adler-Manin residue, [W].

The Radul cocycle in $n$ dimensions is  defined as
$$c(a,b) = Res([log|p|,a]*b).\tag3.7$$
For multiplication operators the Radul cocycle
vanishes in dimensions higher than one. The structure of the Lie group
defined by the Radul cocycle has recently been studied in [KV].

\vskip 0.3in
4. RENORMALIZED CURRENTS AS PSDO's

\vskip 0.3in

Consider a system of quantized fermions in external vector potentials $A$
in $3+1$ space-time dimensions. $A$ takes values in $\gm,$ a finite-dimensional
Lie algebra. We set up a hamiltonian formalism for second quantization: we
consider field (anti)commutation relations at a fixed time $t=0.$

In Schr\"odinger picture the wave functions are functions $\phi(A)$ of the
potential, with values in a fermionic Fock space $\Cal F.$

We use the temporal gauge $A_0=0$ and we consider only time-independent gauge
transformations.

The (free) Dirac field $\psi$ satisfies the CAR algebra
$$\psi^*_{ia}(x)\psi_{jb}(y)+\psi_{jb}(y)\psi^*_{ia}(x) =\delta_{ij}\delta_{ab}
\delta(x-y)\tag4.1$$
where $i,j$ are space-time indices and $a,b$ are internal symmetry indices;
the latter refer to a unitary representation of $\gm.$

The free vacuum is characterized by the property
$$\psi(u)|0>=0=\psi^*(v)|v> \text{ for } u\in \Cal H_+,v\in \Cal H_-\tag4.2$$
where $\Cal H_{\pm}$ are the positive and negative energy subspaces of the one
particle Hilbert space $\Cal H$ and
$$\psi(u)= \int \psi_{ia}(x) u_{ia}(x) d^3 x.$$

In the perturbation theory based on Dyson expansion one writes the scattering
amplitudes in terms of expressions like
$$<0| H_I(t_1) H_I(t_2)\dots H_I(t_n)|0>.\tag4.3$$
These are integrated over the times $t_1 > t_2 \dots >t_n.$ Divergencies for
the scattering amplitudes occur
since the interaction hamiltonian
$$H_I(x)= \psi^*(x) \g_0\g^k T_a \psi(x) A^a_k(x)\tag4.4$$
involves products of field operators at the same point and does not lead
to a well-defined operator distribution in Fock space. We need a
renormalization
even after normal ordering. (In $1+1$ dimensions a normal ordering is
sufficient!)

More precisely, the technical reason for divergencies is the following.
Let $\epsilon$ be the sign of the free
hamiltonian $H_0= \g_0\g^k\nabla_k$ in the one-particle representation.
Then the off-diagonal blocks $[\epsilon, H_I]$ are Hilbert-Schmidt only
when the space-time dimension is at most 2. As discussed in [Ar], the
Hilbert-Schmidt property is both necessary and sufficient for quantization
of operators of the type
$$H_I= \sum X_{nm} a^*_n a_m$$
where the $a_n$'s satisfy the CAR relations
$$a^*_n a_m + a_m a^*_n=\delta_{nm}$$
The $X_{nm}$'s are the matrix elements of $H_I$ in a basis of eigenvectors
of $H_0$ (compactify the physical space to get a discrete basis). Note that
the norm  of the state $H_I |0>$ is given by
$$||H_I |0>||^2= \sum_{E_n>0, E_m<0} |X_{nm}|^2,$$
where $E_n$'s are the energy levels for $H_0.$
The finiteness of this norm is just the Hilbert-Schmidt property for one
of the off-diagonal blocks of the one-particle operator.

The badly behaving interaction hamiltonian is renormalized as follows. For each
$A$ we construct a unitary operator $T_A$ in the
one-particle space such that
$$[\epsilon, T_A^{-1} (H_0+ H_I) T_A]$$
is Hilbert-Schmidt.

The strategy is to obtain $T_A=1+t_{-1}+t_{-2}\dots$ as an expansion in
homogeneous pseudodifferential operators $t_k$ of degree $k=0,-1,-2,\dots.$
We shall consider separately the left and right handed sectors in the space
of 4-component fermions.

Let us consider the following operator acting on two-component fermions:
$$\align T_A&= 1-\frac{i}{4|p|^2}[p,A]-\frac{1}{32|p|^4}[p,A]^2
-\frac18 \left[\frac{\sigma_k}{|p|^2}-2\frac{pp_k}{|p|^4},\partial_k A
\right]\\
&-\frac{1}{8|p|^4}[p,A] (A\cdot p) -\frac{1}{8|p|^4}(A\cdot p)[p,A]
+ O(-3) \tag4.5 \endalign$$
We have used the following notation: $p=p_k\sigma_k,$ $A=A_k\sigma_k$
The commutators in (4.5) are all ordinary commutators of Pauli matrices
$\sigma_k$ and of Lie algebra elements in $\gm.$ The Pauli matrices are
normalized such that $\sigma_1\sigma_2=i\sigma_3$ and $\sigma_k^2=1.$

Since the physical space is assumed to be compact, there can be no
infrared divergencies in the theory, so it sufficient to work with the
asymptotic expansions of the operators in momentum space; the asymptotic
expansion gives a complete picture of the ultraviolet properties of the
theory. If one feels uneasy about the 'infrared singularity' at $p=0$ of
the terms in the asymptotic expansion one can always replace the inverse
powers $1/|p|^n$ by some smooth functions which agree with the original
for large values of $|p|.$

We can write $T_A^{-1} (H_0+H_I) T_A=H_0 +W_A.$
After a tedious computation we obtain
$$\align W_A&=T_A^*(p+iA)T_A -p=\frac{ip}{|p|^2} A\cdot p \\
&-\frac18\left[p,[\frac{\sigma_k}{|p|^2} -2\frac{pp_k}{|p|^4},
\partial_k A]\right] -\frac{\sigma_k}{4|p|^2}[p,\partial_k A] \\
&+\frac{i}{2|p|^2}\epsilon_{ljk}
p_j[A_l,A_k]-\frac{p}{|p|^2} A_m A_m +\frac{p}{|p|^4} (A\cdot p)^2
+ O(-2).\tag4.6\endalign$$
It is then a simple computation to show that $[\epsilon,W_A]$ is of degree
$-2,$
and therefore the operator is Hilbert-Schmidt.
There is no magic in the derivation of the formula (4.5) for $T_A.$ It is a
simple
recursive procedure. Writing
$$T_A= 1+t_{-1} + t_{-2} +\dots$$
in the asymptotic expansion, one gets
$$T_A^* (p+ \a_0 +\a_{-1}+ \dots) T_A= p + \a_0' +\a'_{-1}+\dots,$$
where
$$\align \a'_0&= \a_0 +[p,t_{-1}]\\
  \a'_{-1}&= \a_{-1}+[\a_0,t_{-1}]+p t_{-2}+(T^*_A)_{-2}p -i\sigma_k\partial
  _k t_{-1}.\tag4.7\endalign$$
Again, the commutators above are ordinary matrix commutators (and not *-
commutators).

The Hilbert-Schmidt condition on $[\epsilon,W_A]$ is equivalent to the pair of
equations
$$[\epsilon,\a'_0]=0 \text{ and } [\epsilon,\a'_{-1}]-i(\partial_{p_k}
\epsilon) (\partial_{x_k}\a'_0) =0.$$
Inserting from (4.7), the first equation is just a linear algebraic equation
for the symbol $t_{-1}.$ But then we can solve $t_{-2}$ from the second
equation
above.

In the one-particle representation the infinitesimal gauge transformations $X$
are acting on Schr\"odinger wave functions $\phi(A)$ as the operators
$\Cal L_X +X,$ where the first part
(Lie derivative) is the gauge action on vector potential $A$ and the second is
a multiplication operator acting on the value $\phi(A)\in\Cal H.$ After the
conjugation by the $A$
dependent operator $T_A$ these become $\Cal L_X+\theta(X;A)$ with
$$\theta(X;A)= T^{-1}_A X T_A +T_A^{-1}(\Cal L_X T_A).\tag4.8$$
By construction,
$$\align [\theta(X;A)+\Cal L_X,&\theta(Y;A)+\Cal L_Y]
\\&= [\theta(X;A),\theta(Y;A)]
+\Cal L_X \theta(Y;A) -\Cal L_Y\theta(X;A)+[\Cal L_X,\Cal L_Y]
\\&= \theta([X,Y];A)+\Cal L_{[X,Y]}\tag4.9\endalign$$
That is, the functions $\theta(X;\cdot)$ form a 1-cocycle for the gauge
action of $Map(M,\gm).$ If the function $T_A$ is constructed as above, then
$\theta(X;A)$ is in the Lie algebra of the \it restricted unitary group \rm
$U_{res}, [M1].$ The latter is the subgroup of $U(\Cal H)$ consisting of
operators $g$ such that $[\epsilon,g]$ is Hilbert-Schmidt.

\bf Proof: \rm Since $H'(A)= T_A^{-1}(H_0+H_I(A))T_A =H_0 +W_A$ and $[\epsilon,
W_A]$ is
HS, we observe that the sign $\epsilon'(A)$ of the Hamiltonian $H'(A)$ differs
from $\epsilon$ by a HS operator. Let $g$ be a finite gauge transformation.
It acts on Schr\"odinger wave functions by
$$(R(g)\phi)(A)= g\cdot \phi(g^{-1}Ag +g^{-1}dg)$$
But after the conjugation by $T_A:$
$$(R'(g)\phi)(A)= (T^{-1}_{g\cdot A} g T_A)\phi(g^{-1}\cdot A)\equiv
\o(g;A)\phi
(g^{-1}\cdot A).$$
Thus
$$\align [\epsilon, \o(g;A)]&= \epsilon T^{-1}_{g\cdot A} g T_A- T^{-1}_{g
\cdot A} g T_A\epsilon
\\&= (T^{-1}_{g\cdot A}\epsilon T_{g\cdot A}) g T_A
-T^{-1}_{g\cdot A} g(T_A \epsilon T_A^{-1}) T_A
\\&
=T^{-1}_{g\cdot A}\left( (T_{g\cdot A}\epsilon T^{-1}_{g\cdot A})g
-g(T_A \epsilon T^{-1}_A)\right) T_A\\
&\equiv T^{-1}_{g\cdot A}\left(\epsilon(g\cdot A)g-g\epsilon(A)\right)T_A
\endalign$$
where we have used the equivariantness of the family of Dirac operators,
$$g H(A) g^{-1} = H(g\cdot A).$$
Thus finite gauge transformations satisfy the HS condition; considering one-
parameter subgroups one proves the HS condition for the generators.

The asymptotic expansion for (4.8) is
$$\theta(X;A) = X +\frac{i}{4} \frac{[p,dX]}{|p|^2} +\theta_{-2}
+O(-3) \tag4.10$$
with
$$\align \theta_{-2} =&-\frac14 \frac{[\sigma_k,A]}{|p|^2}\partial_k X
                +\frac12 \frac{[p,A]}{|p|^4}p_k\partial_k X\\
               & +\frac{1}{16}\frac{[p,A]}{|p|^4} [p,dX].
               \endalign$$

If $X$ is any bounded bilinear quantity in the fermion creation and
annihilation
operators such that its off-diagonal blocks in the one-particle representation
with respect to the energy polarization are HS, then the second quantized
operator $\hat X$ is well-defined (after normal ordering) and
$$[\hat X,\hat Y]= \widehat{[X,Y]} + c(X,Y),\tag4.11$$
where $c$ is a Schwinger term, [L],
$$c(Y,Y) = \frac14 \TR \epsilon [\epsilon,X] [\epsilon,Y].\tag4.12$$

\bf Example \rm Multiplication operators in $1+1$ dimensions. Multiplication
operators are PSDO's $X$ of order zero with a $p$ independent symbol $a(x).$
$\epsilon= \frac{p}{|p|}$ and
$$[\epsilon, X] = -2i\delta(p) a'(x).\tag4.13$$
Now the trace (4.12) can be written as the conditionally converging trace
$$c(X,Y)= \frac12 X[\epsilon,Y] =\frac{1}{2\pi i} \int \TR\, a(x) b'(x)dx.
\tag4.14$$
This is the central term of an affine algebra.

In the $3+1$ dimensional case one just inserts from (4.8) to (4.12):
$$c(X,Y;A)=\frac14 \TR\epsilon[\epsilon,\theta(X;A)][\epsilon,\theta(Y;A)].
\tag4.15$$
Actually, we can write
$$c(X,Y;A)= \frac12 \TR [\epsilon,\theta(X;A)] \theta(Y;A)\tag4.16$$
as a conditionally convergent trace: Compute first the traces for the finite-
dimensional matrices, perform the momentum space integration over the spherical
angles, next the integration over $|p|,$
and finally integrate the star product of symbols in configuration space.

\vskip 0.3in
5. COMPUTATION OF THE COMMUTATOR ANOMALY

\vskip 0.3in
The result of the computation starting from (4.16) is rather complicated
expression involving terms of all orders in
$A.$ However, there is a great simplification if we are interested only on the
cohomology class of the cocycle. A change in the renormalization $T_A\mapsto
T'_A=T_A g_A$ of the gauge currents (with $g_A \in U_{res}$) leads to a
modification
$$\theta(X;A) \mapsto \theta'(X;A)=g_A^{-1}\theta(X;A)g_A + g_A^{-1}
\Cal L_X g_A.$$
It is easy to check that
$$c' -c = (\delta \l)(X,Y;A)=\l([X,Y];A) -\Cal L_X\l(Y;A) +\Cal L_Y
\l(X;A),$$
where $\l$ is the 1-cochain
$$\l(X;A) = \frac14 \TR \left[\epsilon, \tilde g_A^{-1} [\theta(X;A),\tilde
g_A]
 + \tilde g_A^{-1} \Cal L_X \tilde g_A \right]_+,$$
with $[A,B]_+ =AB+BA,$ and $\tilde g_A =g_A f_A$ for any smooth function
$f_A$ such that 1) the diagonal blocks of $\tilde g_A$ are trace-class, 2)
the off-diagonal blocks of $f_A$ vanish. Such a choice is possible because
the diagonal blocks of each $g_A$ are Fredholm operators and because the
parameter space $\Cal A$ is flat.

We want to show that if we choose $\l$ in a suitable way then the new cocycle
$c'$
takes the form
$$c'(X,Y;A)= \frac{i}{24\pi^2}\int \TR A [dX,dY].$$
Let us first define the regularized trace, with $\L >0,$
$$\TR_{\L} R = \frac{1}{(2\pi)^3}\int_{|p|\leq \L} \TR \,r(x,p) d^3x d^3p
+\frac{1}{(2\pi)^3}\int_{|p|>\L} \TR \left( r - \sum_{k=0}^3 r_{-k} \right)
d^3x d^3p$$
for a PSDO $R$ of degree zero. Define
$$\l(X;A) = \frac12\TR_{\L}\,\epsilon\, \theta(X;A).$$
Then
$$(\delta\l)(X,Y;A) = \frac12 \TR_{\L}\epsilon\,[\theta(X;A),\theta(Y;A)].$$
If the regularize trace were symmetric, this would be equal to the cocycle $c.$
In order to show that $\TR (a*b) =\TR (b*a)$ for a pair of symbols one has to
perform
partial integration both in the momentum and coordinate variables. The
coordinate
integration does not cause any problems, since we assumed that the manifold is
compact and without boundary. However, there can be boundary terms arising from
the momentum space integration. Integrating a term of degree $-3$ leads to
finite boundary contribution in partial integration: the integrand
behaves like $|p|^{-2}$ at infinity, cancelling the factor $|p|^2$ coming
from the integration measure. The boundary contributions from terms of degree
less than 3 vanish at $|p|\to\infty.$

As we have seen, the terms in $c$ which are not coboundaries of some 1-cochains
must be of order $-3.$ The terms of order greater than -3 vanish in (4.15) by
the
algebra of Pauli matrices and by the fact that the momentum space integration
over spherical angles of a symbol of odd degree gives zero.  According to the
multiplication rule (3.5), with $X_k\equiv \theta(X;A)_k,$
$$\align S\equiv \left(\epsilon[\theta(X;A),\theta(Y;A)]\right)_{-3}& =
\epsilon\left([X,Y_{-3}] + [X_{-1}, Y_{-2}] \right. \\
& \left.+\{X,Y_{-2}\} + \{X_{-1},Y_{-1}\} +\{X_{-2},Y\} - (X \leftrightarrow Y)
\right),\endalign$$
where we have denoted $\{a,b\}= -i \sum\partial_{p_k} a \partial_{x_k} b,$
'half Poisson bracket'. The commutators on the right are matrix commutators.
The cutt-off trace of the $\{X_{-1},Y_{-1}\}$ term is seen to vanish after
performing
the spherical part of the momentum space integration. By partial integration
we obtain
$$\align \TR_{\L} S &= \TR_{\L} \left( [\epsilon, X_{-1}]Y_{-2} +
[\epsilon,X_{-2}]Y_{-1} - \{\epsilon,Y\}X_{-2} +
\{\epsilon, X\}Y_{-2}\right)\\
&+\frac{i}{(2\pi)^3} \int_{|p|=1} \TR\, \epsilon Y_{-2} (p_k \partial_{x_k}
X) \eta (d\eta)^2 -\frac{i}{(2\pi)^3} \int_{|p|=1} \TR\,
\epsilon X_{-2} (p_k\partial_{x_k} Y)\eta (d\eta)^2 \\
&= \frac12 \TR_{\L}\left(\left( [\epsilon,\theta(X;A)]*\theta(Y;A)\right)_{-3}
-\left( [\epsilon,\theta(Y;A)]*\theta(X;A)\right)_{-3}\right) \\
& + Res\,\, \epsilon [log(|p|),\theta(X;A)]*\theta(Y;A).\tag5.1\endalign$$
It follows that
$$c_{\L}(X,Y;A) = Res\,\, \epsilon [log(|p|),\theta(X;A)]\theta(Y;A).\tag5.2$$
(Here star product shoud be used). The residue is easely computed (as the
spherical integrals in (5.1)). The result
is
$$c_{\L}(X,Y;A) = \frac{i}{24\pi^2} \int_x \TR\, A[dX,dY],\tag5.3$$
which is the cocycle derived in [M, F-Sh] in a different context. Note that
the final results (5.2) and (5.3) do not depend on the cut-off parameter
$\L.$ This method of computing cocycles can be generalized to various
directions, [LM].

The action functional in Connes noncommutative geometry model of Yang-Mills
theory is also defined in terms of the Wodzicki residue, [C]. It would be
interesting
to see more precisely the relation between the present hamiltonian approach and
the Lagrangian method of Connes.

Actually the formula (5.2) can  be applied to the algebra $W'$ of all PSDO's
$P$
(in three dimensions) obeying the condition
$$deg\,[\epsilon,P] \leq -2\tag5.4$$ giving
a twisted form of the Radul cocycle on the Lie algebra $W'$,
$$c(P, Q) = Res \,\,\epsilon [log(|p|), P] Q \tag5.5$$
So we get a generalization of the $W_{\infty}$ algebra  as the central
extension (defined by the 2-cocycle (5.5)) of the Lie algebra of
restricted spinorial PSDO's in three  dimensions. In $n$ dimensions the
condition (5.4) should be replaced by the requirement that the degree of the
commutator is strictly less than $-\frac12 n.$  We shall end the discussion
by giving the proof that (5.5) is a 2-cocycle for the restricted Lie algebra
$W'$ is $n$ dimensions. Denote
$$Res' A = Res \,\epsilon A.$$
Since $Res[A,B]=0$ for any pair of PSDO's $A,B$ we have
$$\align Res\,\epsilon[\epsilon,A][\epsilon, B]&= Res(A\epsilon B -\epsilon AB
-AB\epsilon +\epsilon A\epsilon B\epsilon)\\
&= 2Res(A\epsilon B -\epsilon AB)
=-2 Res[\epsilon A]B=-2 Res\epsilon[A,B].\endalign$$
If now $deg[\epsilon, A]< -n/2$ and $deg[\epsilon, B]< -n/2$ then $deg\epsilon
[\epsilon,A][\epsilon,B] < -n$ and it
follows that the residue vanishes. Thus
$$Res'[A,B]=0\tag5.6$$
for any $A,B\in W'.$ Denote $\ell= log(|p|).$ By (3.5) and performing partial
integration in momentum space one gets $Res[\ell,A]=0$ for any PSDO $A.$
Since $\ell$ commutes with $\epsilon,$ we have also
$$Res'[\ell,A]=0\tag5.7$$
although $\ell$ is not a PSDO. Define
$$\o(A,B,C)= c(A,[B,C]) +c(B,[A,C]) + c(C,[A,B]).$$
Then
$$\o(A,B,C)= Res'([\ell,A][B,C]+[\ell,B][C,A]+[\ell,C][A,B]).\tag5.8$$
Using (5.6) and (5.7) we get
$$\o(A,B,C)= Res'\left(A[[B,C],\ell]+ A[[\ell,B],C]+A[[\ell,C],B]\right)=0,$$
by Jacobi's identity, proving that  $c$ is indeed a 2-cocycle.

The formula (5.3) can be generalized to higher (odd) space dimensions in
various ways. One approach is to use the cohomological descent equations
starting from a Chern class in dimension $n+3,$ [M, F-Sh], [M2]. There
exists another rather natural and much simpler generalization, which was found
recently
when studying p-brane symmetries, [CFNW]. The realization of that algebra
as an algebra of PSDO's (of type $W'$) is now under preparation.

\vskip 0.3in
REFERENCES  \newline\newline

[Ad] M. Adler, Invent. Math. \bf 50, \rm 219 (1979)

[Ar] H. Araki in: \it Contemporary Mathematics, \rm vol. 62, American
Mathematical Society, Providence (1987)

[C] A. Connes, Commun. Math. Phys. \bf 117, \rm 673 (1988)

[CFNW] M. Cederwall, G. Ferretti, B. Nilsson, and A. Westerberg, preprint
ITP 93-37 Gothenburg; HEP-TH/940127

[H] L. H\"ormander: \it The Analysis of Linear Partial Differential
Operators III. \rm Springer-Verlag, Berlin (1985)

[KV] M. Kontsevich and S. Vishik, preprint Max-Planck-Institut f\"ur
Mathematik, Bonn
(April 1994); HEPTH 94 04 46

[KK] O.S. Kravchenko and B.A. Khesin, Funct. Anal. Appl. \bf 25, \rm
83 (1991)

[L] Lars-Erik Lundberg, Commun. Math. Phys. \bf 50, \rm 103 (1976)

[LM] E. Langmann and J. Mickelsson, work under preparation

[Ma] Yu.I. Manin, J. Sov. Mat. \bf 11, \rm 1 (1979)

[M1] J. Mickelsson, Lett. Math. Phys. \bf 28, \rm 97 (1993)

[M2] J. Mickelsson: \it Current Algebras and Groups. \rm Plenum Press,
London (1989)

[M,F-Sh] J. Mickelsson, Commun. Math. Phys. \bf 97, \rm 361 (1985);
L. Faddeev and S. Shatasvili, Theoret. Math. Phys. \bf 60, \rm
770 (1984)

[R] A.O. Radul, Funct. Anal. Appl. \bf 25, \rm 25 (1991)

[W] M. Wodzicki: Noncommutative residue. In Lecture Notes in Mathematics
1289, ed. by Yu.I. Manin, Springer-Verlag

\enddocument